\documentclass[twocolumn,prl]{revtex4}

\usepackage{graphicx}
\usepackage{amsmath}

\begin{document}

\title{A possible path to a new class of ferromagnetic and half metallic 
ferromagnetic materials}

\author{I.S.~Elfimov$^1$, S.~Yunoki$^1$ and G. A.~Sawatzky$^{1,2}$}

\affiliation{
$^1$Solid State Physics Laboratory, Materials Science Center, \\
University of Groningen, Nijenborgh 4, 9747 AG Groningen, The Netherlands \\
$^2$Department of Physics and Astronomy, University of British Columbia, \\
6224 Agricultural Road, Vancouver, B.C. V6T 1Z1, Canada}

\date{\today}

\begin{abstract}
We introduce a path to a possibly new class of magnetic materials whose
properties are determined entirely by the presence of a low concentration of
specific point defects in their crystal structure. Using model  Hamiltonian and
{\it ab-initio} band structure methods we demonstrate that even large band gap
nonmagnetic materials as simple as CaO can exhibit extraordinary properties
like half metallic ferromagnetism upon introducing a small concentration of Ca
vacancies. We show that such defects will initially bind the introduced charge
carriers at neighboring sites and depending on the internal symmetry of the
clusters formed by neighboring sites form "local" magnetic moments which for
concentrations as low as 3\% transform  this non-magnetic insulator into a half
metallic ferromagnet.
\end{abstract}

\pacs{71.55.-i, 73.22.-f, 75.10.-b, 75.75.+a}

\maketitle

The recent discovery of ferromagnetism with high transition temperatures and
very small magnetic moments in the hexaborides \cite{Young99}, Co substituted
TiO$_2$ \cite{Matsumoto01}, and Co substituted ZnO \cite{Ueda01}
has opened a lively  discussion regarding the role of defects in transforming
insulating non magnetic compounds into ferromagnets with interesting properties
In the hexaborides Monnier and Delley demonstrated, using  density functional
band structure approaches, that neutral B$_6$ vacancy's in a superlattice can
result in a ferromagnetic ground state even for a low density of such defects
\cite{Monnier01}. We note that the nominal valence of the B$_6$ clusters in
fairly ionic materials such as CaB$_6$ is $2-$ and so two electrons are
needed to charge compensate for such a B6 vacancy.
It is also interesting to note that substituting
divalent Co for a tetravalent Ti in TiO$_2$ also can be compensated with two
holes in the Oxygen 2$p$ valence bands which could be bound impurity states in
close proximity to the Co impurity. These observations may remind some that
in hydrocarbon molecules with  ring structures Longuet-Higgins  predicted
magnetic ground states due to a kind of molecular Hund's rule coupling well
known in atoms \cite{Longuet50}. This property has in fact been used
in the past by Torrance {\it et al} \cite{Torrance88} in attempts to
make organic molecular ferromagnets. More recently Eskes has used a Hubbard
model calculation of ring systems and clusters to demonstrate again a
kind of molecular Hund's rule coupling for orbitally degenerate states, 
leading to very stable magnetic (spin triplet) ground states for either two 
electrons or two holes, in particular  geometries \cite{Eskes}.

\begin{figure}
\includegraphics[clip=true,width=0.49\textwidth]{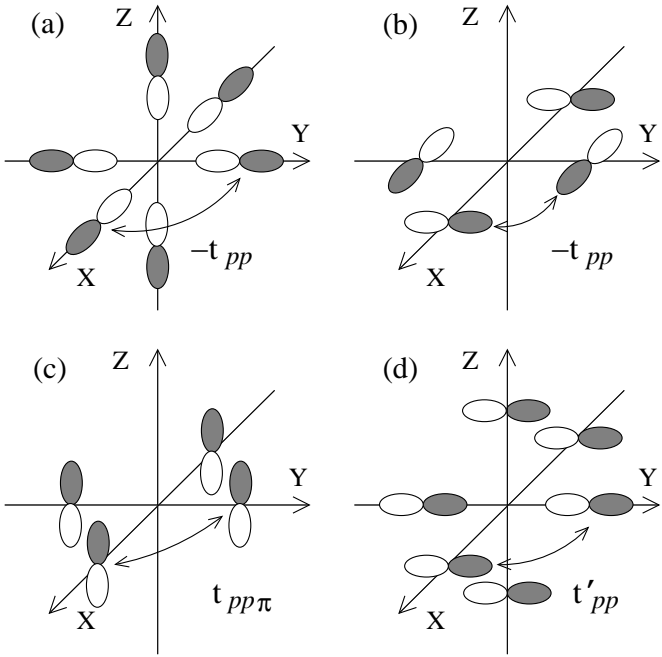}
\caption{
An artists concept of the oxygen $\sigma$ (a) and $\pi$ (b,c) bonding orbitals
relative to the O-vacancy bond direction surrounding a Ca vacancy. 
Also shown are the definitions of the hopping integrals (b) $t_{pp}$,
(c) $t_{pp\pi}$, and (d) $t'_{pp}$ given in terms of electrons.
}
\label{fig:hoping}
\end{figure}

In the present work we report on the influence of dilute divalent cation
vacancies in oxides with the rock salt structure. We predict that such
systems have ferromagnetic ground states with small magnetic moments but
possibly high transition temperatures. For 3.125\% Ca vacancies in CaO LDA band
structure calculations predict a half metallic ferromagnetic ground state
which if correct could play an important role in the modern field of
spintronics and opens a path to a whole new class of ferromagnetic materials.
Before presenting the band structure calculations we discuss the
physics we believe to be responsible for the magnetic behaviour found. In the
rock salt structure the cations are surrounded by an octahedron of anions with
filled valence $p$ bands as shown in Fig.\ref{fig:hoping}.
As also shown, the $\sigma$-bonding anion $p$ orbitals with lobes directed to
the cation site, would be the most energetically favourable orbitals to
accommodate the holes needed to charge compensate a cation vacancy at the
center of such an octahedron. Because of the negative effective charge of the
cation vacancy the holes in the anion valence $p$ band feel a strong attraction
to the vacancy site and if large enough this attraction can bind two holes
as in fact is found in the LDA calculations.

In order to get insight into the electronic structure of such an
octaheral cluster of O$^{2-}$ ions each with a full 2$p$ shell we use a tight
binding like molecular orbital (MO) approach. The symmetries of the MO's
formed by the $\sigma$-bonding orbitals are $a_{1g}$, $t_{1u}$, and $e_g$
at energies equal to $-4t_{pp}+\epsilon_\sigma$, $\epsilon_\sigma$,
$2t_{pp}+\epsilon_\sigma$ respectively. Note that simple
sum of orbitals shown in Fig.\ref{fig:hoping}(a) represents the
$a_{1g}$ symmetry state and this state has the lowest energy for electrons
and the highest for holes. The $\pi$-bonding orbitals form the symmetries
$t_{2g}$, $t_{1u}$, $t_{2u}$, and $t_{1g}$ and have energies  $-2t_{pp}$,
$2t_{pp\pi}$, $-2t_{pp\pi}$, and $2t_{pp}$ respectively where
$t_{pp}=1/2(t_{pp\sigma}-t_{pp\pi})$ with the $\sigma$ and $\pi$ hopping
integrals now relative to the O-O bond direction and given in terms of
electrons ($t_{pp\pi}$ is negative and
$t_{pp\sigma} \sim -4t_{pp\pi}$). Note that for simplicity the zero of energy
is chosen to be at $\epsilon_\pi$. One can see that the irreducible
representation $t_{1u}$ appears in both the $\sigma$ and $\pi$ combinations.
As a result the molecular orbitals of only this particular symmetry will mix
with each other resulting in a shift of their energies to
$1/2\epsilon_\sigma + t_{pp\pi} \pm
1/2\sqrt{(\epsilon_\sigma-2t_{pp\pi})^2+32t^{'2}_{pp}}$
for the MO formed by the $\sigma$ ("$+$") and $\pi$ ("$-$") bonding orbitals.
The hopping integral $t'_{pp}$ is defined as
$1/2(t_{pp\sigma}+t_{pp\pi})$ and is shown in Fig. \ref{fig:hoping}(d).

\begin{figure}
\includegraphics[clip=true,width=0.47\textwidth]{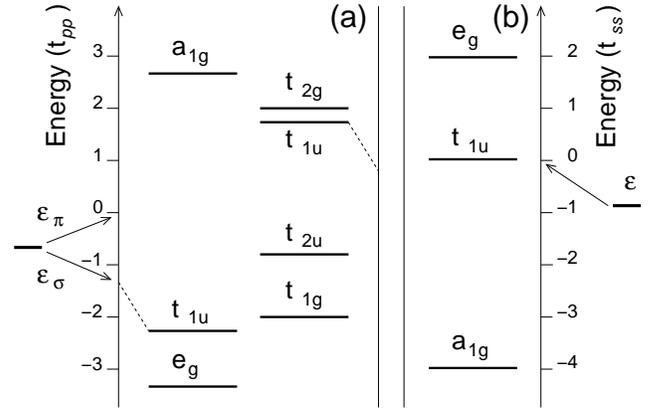}
\caption{
Schematic single particle energy level diagram for (a) HOLES in anion
orbitals and (b) ELECTRONS in cation orbitals. Dash line in fig. (a)
represents a shift of energy levels of molecular orbitals with $t_{1u}$
symmetry due to the mixture between $\sigma$ and $\pi$-bonding anion orbitals.
The energy of the atomic $\sigma$ and $\pi$ orbitals are indicated by
arrows on the left hand margine.
}
\label{fig:energy_single}
\end{figure}

The schematic energy diagram is given in Fig.~\ref{fig:energy_single}.
The lowest energy states for holes are the $e_g$ states formed by a linear
combination of the $p$ orbitals with lobes directed to the vacancy with phases
such as to result in a $e_g$ symmetry state when viewed from the vacancy
and the $t_{1g}$ states with a corresponding linear combination of $p$
orbitals with lobes directed perpendicular to the O-Vacancy bond direction.
A simple calculation shows that a hole in $p$
orbitals with lobes pointing towards the vacancy has an energy about 1eV
lower than a hole in a $p$-orbital with lobes perpendicular to the O-Vacancy
bond direction because of the crystal field produced by the effective $2-$
charge of the cation vacancy. Therefore the lowest energy  single hole state
will be the $doubly$ $degenerate$ $e_g$ molecular orbital
(Fig.~\ref{fig:energy_single}(a)).
We note that for anion vacancies the conduction band orbitals of the cation
would be occupied by the charge compensating electrons and their energy would
be lowest for the non-degenerate $a_{1g}$ orbital as also shown in
Fig.\ref{fig:energy_single}(b).

\begin{figure}
\includegraphics[clip=true,width=0.4\textwidth,angle=-90]{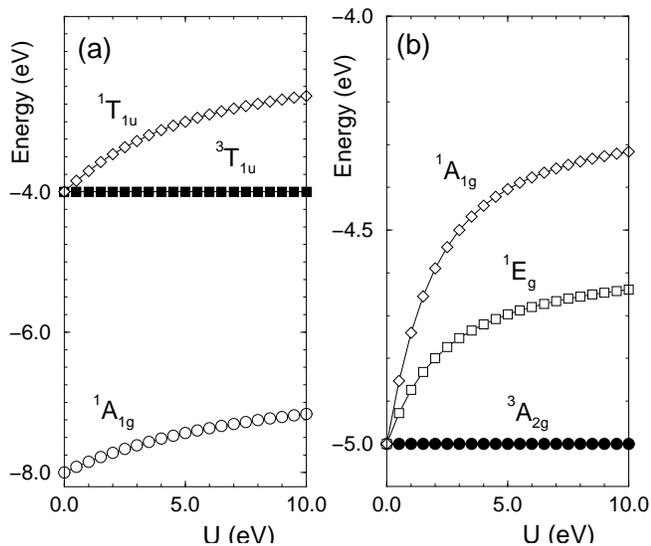}
\caption{ Energy diagrams of the lowest three states for two
particles in an octahedral cluster as a function of the on-site
Coulomb repulsion $U$; (a) Two electrons
introduced into the empty $s$-orbitals with
the hopping integral
$t_{ss}=-1.0$ eV and (b) two holes introduced into the completely
filled $p$-orbitals with the hopping integrals $t_{pp\sigma}=1.2$ eV
and $t_{pp\pi}=-0.3$ eV and the crystal field potential
$\varepsilon_\sigma=1$ eV and $\varepsilon_\pi=0$ eV given in terms of
electrons. Solid (open) symbols are for spin triplet (singlet) states.}
\label{fig:e_of_2_particles}
\end{figure}

Having understood the single-particle molecular orbitals, we move to the
case of two holes (electrons) in the completely filled (empty) shell formed
by the anion $p$ (cation $s$)  orbitals in the case of cation (anion) vacancies.
If the Coulomb interaction between those particles is negligible, then two holes
will occupy the $e_g$ molecular orbital and they will form three degenerate
states $^1A_{1g}$, $^1E_g$ and $^3A_{2g}$ (Fig.\ref{fig:e_of_2_particles}(b)).
Two electrons on the other hand will result in a non degenerate singlet ground
state (Fig.\ref{fig:e_of_2_particles}(a)). Let us now switch on the
Coulomb repulsion of which the dominant term will be that when two particles
are on the same site as in a Hubbard model. The three states found for the
configuration with two holes will now split up because for the singlet states
the two holes can be in the same site and in the same $\sigma$ orbital while
this is forbidden for the spin triplet state. This means that the singlets will
increase in energy leaving us with a triplet and magnetic  ground state. From
Auger spectroscopy studies of a large number of simple oxides the on-site
cooulomb repulsion energy of two holes in an oxygen $p$ orbital is 5-7 eV
\cite{Tjeng88,Altieri}.

To shed more light on this issue we carried out exact diagonalization
calculations for
the single octahedral cluster with two electrons in  $s$ orbitals as well
as with two holes in the $p$ orbitals of a octahedral cluster taking
into account the on-site Coulomb interaction $U$ exactly, and the results
are shown in Fig.~\ref{fig:e_of_2_particles}. One can immediately see that
while a spin singlet state is the ground state for two $electrons$ independent
of the value of $U$, a spin triplet state is the ground state for the two
$holes$ as soon as $U$ is different from zero. This triplet state has symmetry
$^3A_{2g}$ and the wave function consists mainly of a $(e_{g})^2$
configuration shown schematically in Fig.~\ref{fig:energy_single}(a).
Note that the lowest singlet state is at about 0.3--0.4 eV higher in energy
than the triplet ground state of the two holes for realistic parameters. As
mentioned above this is a kind of molecular Hund's rule coupling and is a
result of the degenerate nature of the lowest energy molecular orbital for
holes in a octahedral cluster. We note that for  oxygen vacancy's
it does not matter if we are dealing with $s$ or $p$, $d$ states of the cations.
As long as band is empty the lowest energy state for electrons
in octahedral geometry will always be a totally symmetric state and therefore
non degenerate.

Now that we have found that cation vacancies in simple divalent monoxides with
the rock salt structure or in general cation vacancies in compounds with
octahedral coordination form impurity states with local magnetic moments we may
ask what the magnetic state of the material is for a finite concentration of
such vacancies. Do they couple ferromagnetically or antiferromagnetically and
is the resulting material metallic or still insulating? To address this we
consider the problem of a super lattice of cation vacancies with a 2x2x2 
super-cell. This would correspond to a vacancy concentration of only 3.125\%. 
We calculate the electronic structure using a density functional band structure
code (TBLMTO-47 \cite{lmto}) based on the local density approximation (LDA). For
the Ca atoms, the basis from 4$s$, 4$p$, and 3$d$ orbitals was used, while for
oxygen, the 2$s$, 2$p$ and 3$d$ states were considered. The electronic states
inside the empty sphere between the atoms were expanded up to 3$d$ states
including 1$s$ and 2$p$ orbitals.

Before we discuss the results of band structure calculations we note that,
due to the ionic character of this compound, the valence band is formed of
mainly oxygen 2$p$ orbitals and is full and the conduction band is formed of
mainly Ca orbitals and is empty in the vacancy free compound.
Pure CaO is therefore found to be a diamagnetic insulator with a calculated
band gap of 3.46 eV. The experimental gap is 7 eV \cite{Whited69}.

\begin{figure}
\includegraphics[clip=true,width=0.47\textwidth]{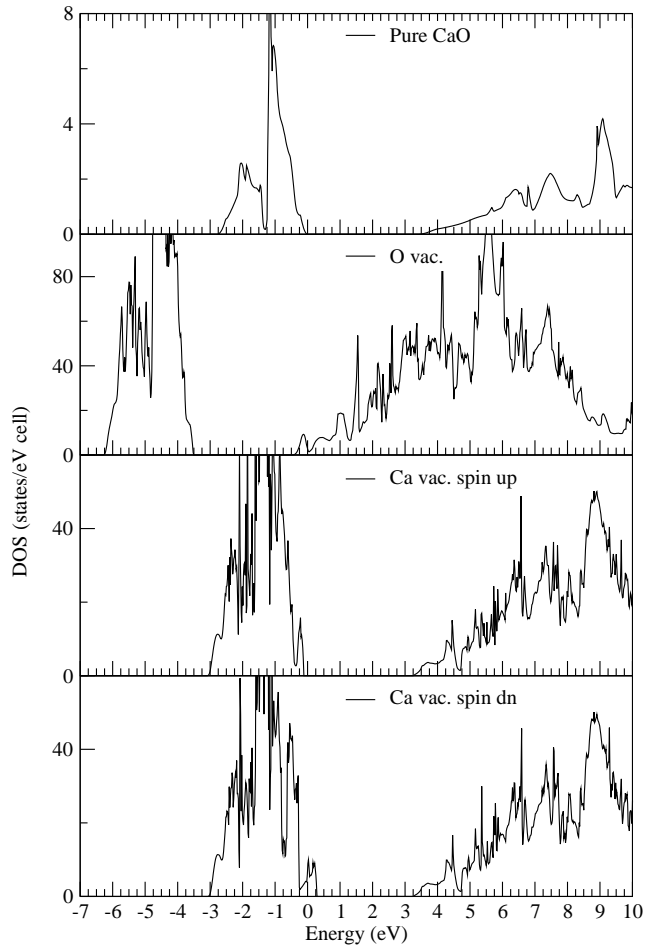}
\caption{Total density of states of pure CaO (upper panel) and
compound with vacancy in anion (second panel) and cation (two
lower panels) sublattices. The zero of energy is at Fermi energy.}
\label{DOStotal}
\end{figure}

In Fig.~\ref{DOStotal} we display the total density of states of pure CaO and
CaO with vacancies in the cation or anion sublatticies. We observe the impurity
state just below the conduction band and the chemical potential shifted up
above this impurity state for the case of Oxygen vacancies. This material is a
non magnetic metal. The same calculation for MgO
with Oxygen vacancies exhibits a clear gap separating the defect induced band
from the conduction band resulting in an insulating non magnetic ground state.
The details of calculations on MgO, CaO, and SrO will be presented elsewhere.

In sharp contrast the band structure for Ca vacancies exhibits also well defined
defect state bands crossing the Fermi energy resulting in a metallic solution.
Also here we see the clear split off character of the vacancy induced states
consistent with the physical picture we presented above of the effects of the
defect potential. Even more interesting is the fact that the Ca vacancy
material is found to be ferromagnetic and the bands crossing the Fermi energy
are totally spin polarized. A Ca deficiency results in a magnetic moment of
2$\mu_B$ per super cell, 88\% of which is concentrated on the six Oxygen ions
that are the nearest neighbors to the Ca vacancy.

\begin{figure}
\includegraphics[clip=true,width=0.47\textwidth]{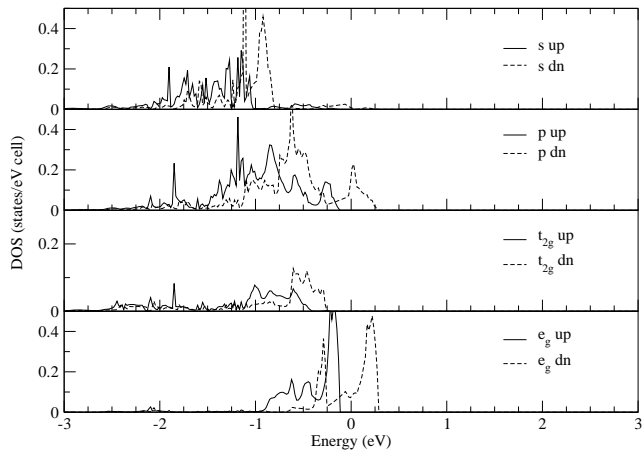}
\caption{Ca vacancy partial density of states. The zero of energy is at Fermi
energy.} \label{CavacDOS}
\end{figure}

\begin{figure}
\includegraphics[clip=true,width=0.47\textwidth]{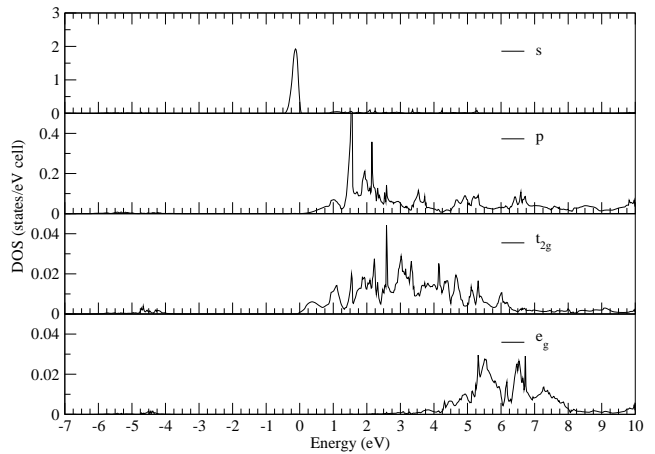}
\caption{Density of states projected
onto the O vacancy site. The zero of energy is at Fermi energy.}
\label{OvacDOS}
\end{figure}

In both cases of Ca or oxygen
vacancies we find that the charge compensating holes or electrons are rather
strongly bound to the vacancy. In Fig.~\ref{CavacDOS} and Fig.~\ref{OvacDOS}
we show the partial density of states projected on the impurity site. It is easy
to see that in the case of Ca vacancies the holes are mainly $d$ like with $e_g$
symmetry but band structure effects also introduce quite a strong component of
$p$ symmetry at the Fermi energy resulting in a half-metallic ferromagnet.
On the other hand, electrons
induced by oxygen vacancies occupy local  molecular orbitals of $s$ symmetry
and result in a nonmagnetic ground state.  Both of these agree very well with
the results of our cluster calculations.
We suggest that the picture presented describes a path to  new classes of
magnetic materials and is a very general phenomenon.

The general underlying rule for obtaining a magnetic ground state is that the
crystal structure must be such that the ground state of the charge compensating
molecular orbital is orbitally degenerate. Secondly the local impurity
potential should be large  enough to quite strongly bind the charge to the
nearest neighbor atoms which is needed to validate the treatment as a cluster
or molecule which then will result in the formation of a "local" moment. An
on-site Coulomb interaction will then result in a high spin ground state for
more than one electron or hole in such a molecular orbital which will mediate a
ferromagnetic magnetic coupling between such clusters via either a
superexchange or a double exchange like mechanism. We note that for the case
considered above the charge compensating
clusters have considerable overlap even though the concentration is only a few
per cent. We also note that such degenerate ground states are common for high
symmetry systems such as ring like structures in addition to the octahedron
considered here.

In conclusion we have demonstrated that point structural deformations of
crystals such as vacancies can indeed confine the compensating charges in
molecular orbitals formed by atomic orbitals on the nearest neighbours. We have
shown that under certain conditions local magnetic moments will be formed due
to a kind of molecular Hunds rule coupling with energetics determined by
kinetic energy and symmetry considerations rather than exchange interactions.
This can lead to high temperature ferromagnetic ground states and in the
example discussed even a half metallic ferromagnetic ground state. Because of
the rather general nature of the physics presented we believe that this may
point to a path towards a new class of ferromagnetic materials.

\end{document}